\documentclass{aa}  
\titlerunning{Evolution of central galaxy alignments in simulations}
\authorrunning{Rodriguez, Merchán \& Artale}
\usepackage{graphicx}
\usepackage{txfonts}
\usepackage{xcolor}
\usepackage{soul}

\usepackage{natbib}
\usepackage[breaklinks=true]{hyperref}% To add links in your PDF file, use the package "hyperref"
\bibpunct{(}{)}{;}{a}{}{,} % to follow the A&A style
\bibstyle{aa} % style aa.bst

\begin{document} 

   \title{Evolution of central galaxy alignments in simulations}

   %\subtitle{Analysis through the anisotropic correlation function and the probability density of the angles }

   \author{Rodriguez, F.
   \inst{1,2}
          \and
          Merchán, M.\inst{1,2}
          \and
          Artale, M. C.\inst{3}
          }

   \institute{CONICET. Instituto de Astronomía Teórica y Experimental (IATE). Laprida 854, Córdoba X5000BGR, Argentina.\\
              \email{facundo.rodriguez@unc.edu.ar}
         \and
             Universidad Nacional de Córdoba (UNC). Observatorio Astronómico de Córdoba (OAC). Laprida 854, Córdoba X5000BGR, Argentina.\\
             \email{manuel.merchan@unc.edu.ar}
        \and
            Universidad Andres Bello, Facultad de Ciencias Exactas, Departamento de Ciencias Físicas, Instituto de Astrofísica, Av. Fernández Concha 700, Santiago, Chile\\
            \email{maria.artale@unab.cl}
            }
            
   \date{Received May 03, 2024 / Accepted May 15, 2024}

% \abstract{}{}{}{}{} 
% 5 {} token are mandatory
 
  \abstract
   {Observations suggest that red central galaxies align closely with their group galaxies and the large-scale environment.This finding was also replicated in simulations, which added information about the alignment of the stars that form the galaxies with the dark matter in the halo they inhabit. These results were obtained for the present universe. Our study aims to build upon previous findings by examining the evolution of central galaxy alignment with their environment, as well as the alignment between their stellar and dark matter components.}
   {Based on previous studies, in this work, we describe the evolution of the alignment of bright central galaxies over time and try to understand the process leading to the current observed alignment.}
   {By employing the merger trees from the simulation, we track the alignment evolution of the central galaxy sample at z=0 used in a previous study, whose results correspond to the observations. In particular, we exploit the anisotropic correlation function to study the alignment of the central galaxies with the environment and the probability distribution of the angle between the axes of the shape tensor calculated for each component to deepen the analysis of the stellar and dark matter components.}
   {A description was given of the evolution of alignment in bright central galaxies with a focus on the distinctions between red and blue galaxies. Furthermore, it was found that the alignment of the dark matter halo differs from that of the stellar material within it. According to the findings, the assembly process and mergers influenced the evolution of alignment.}
   {}


   \maketitle
%
%-------------------------------------------------------------------

\section{Introduction}

 It is well established that the way galaxies form and evolve is linked with the dark matter halo properties they inhabit. 
 The shape, angular momentum and alignment of dark matter halos are  consequence of accretion and tidal processes with the surrounding matter. In this scenario, galaxies are expected to have preferred shapes, orientations and distributions depending on the halo in which they form \citep[see, e.g.,][]{Ciotti1994, Usami1997,Pen2000,Catelan2001,Crittenden2001,Jing2002,Porciani2002a, Porciani2002b,Fleck2003}. 
 Previous studies provide further understanding on how the environmental conditions, in combination with baryonic processes such as stellar and AGN feedback, influence galaxies and, at the same time, use galaxies as proxies for the properties of the halos they reside in. 

Observational results provide insightful information on how galaxies align with each other and with the surrounding structures on large scales. It has been shown that the alignments depend on the luminosity, colour, and star formation history of the galaxies, but also the position within the dark matter halo. In particular, on various scales, red galaxies tend to be more aligned than blue or late galaxies. This would be a consequence of differences in their assembly histories \citep[see, e.g.,][]{Sales2004,yang2006alignment,Agustsson2010,Kirk2015, Rodriguez2022, Smith2023, Desai2022}.
This evidence is also noticeable when analyzing the alignment of satellite galaxies with the central galaxies in their groups. Red satellite galaxies show a stronger preference for aligning with the galactic plane of red central galaxies \citep{Yang2005,Wang2008,Kiessling2015,Kirk2015,Libeskind2015,Welker2018,Pawlowski2018,Johnston2019}. Some authors refer to this effect as \textit{anisotropic quenching} or \textit{angular conformity} \citep{Wang2008,Stott2022,Ando2023}, linked to galactic conformity \citep[see][and references therein]{Bray2016,Otter2020,Maier2022,lacerna2022}.

In \cite{Rodriguez2022}, the alignment of the central galaxies of the groups with the environment is determined using the anisotropic correlation function \citep[see][]{paz2011alignments} calculated using the spectroscopic data provided by Sloan Digital Sky Survey Data Release 16 \citep[SDSS DR16,][]{ahumada2020} and implementing the group finder of galaxies presented in \cite{rodriguez20}. The results show that the bright central galaxies align with both the satellites inhabiting their same halo and the nearby structures of up to more than $\sim$ 10 Mpc. They also found that the alignment shows a strong dependence on colour.

The hydrodynamic cosmological simulations, are an excellent complement to observations and have been fruitful for studying the spatial distribution of galaxies and their intrinsic alignments, as well as that of the dark matter halos that contain them \citep[e.g.][]{Deason2011,Dong2014, Velliscig2015,Shao2016,Welker2018,Zjupa2020,Shi2021,Shi2021,Shi2021b,RagoneFigueroa2020,Samuroff2020,Tenneti2021,Zhang2022, Xu2023}.
In particular, \cite{Rodriguez2023} use IllustrisTNG\footnote{\url{http://www.tng-project.org}} (hereinafter simply TNG), \cite{Rodriguez2023} to study the three-dimensional anisotropic correlation function and  
evaluate the consistency with observational results from \cite{Rodriguez2022}.
Besides finding a general agreement with the main observational results, this work allowed us to deepen the study of galaxy properties and groups not available in observations as well as to extend the analysis. Including the dark matter distribution, a concatenation of physical processes can explain the alignment of galaxies found in the simulations (and observations): the baryonic material of the central galaxy aligns with the dark matter sub-halo it inhabits; the latter aligns with the host halo, and the host halo, in turn, aligns with the large-scale structures.

Based on the excellent agreement between observational results and simulations for redshifts $z=0$ \citealp{Rodriguez2022, Rodriguez2023}, here we study the evolution of the alignment of the central galaxies selected at $z=0$ with the environment to understand how the central galaxies reach their current alignment. For this, we use the same sample as in \cite{Rodriguez2023} and, taking advantage of the data provided by TNG for each snapshot and the merger trees that link them, we calculate the three-dimensional correlation function, as well as the probability distribution of the angles and the triaxiality of the principal axes at different times in the Universe. With this approach, we can comprehend the connection between alignment and dark matter halos and galaxy properties.
 Previous works using IllustrisTNG investigate how the intrinsic alignment evolves with redshift, between $z= 0 - 1$, based on the stellar mass and environmental properties of galaxies \citep{Zjupa2020,Zhang2022,Delgado2023}. In particular, \citet{Zjupa2020} find that elliptical galaxies present a correlation between the tidal field and the observed ellipticity, and this trend is stronger at higher redshifts. On the other hand, spiral galaxies do not present intrinsic alignments as in the case of elliptical galaxies. In this context, our work complements previous findings by studying the evolution of the intrinsic alignment of central galaxies selected at $z=0$ through the anisotropic correlation function.

The paper is structured as follows. In Section ~\ref{data}, we provide a description of the TNG simulation data and the properties that were analyzed. Section ~\ref{evo} presents the results for the anisotropic correlation function for central galaxies split by halo colours, halo mass and the number of mergers. In this section, we also analyze the stellar-dark matter alignment through the angle between the axes of the shape tensor using both stars and dark matter, as well as triaxiality. Finally, we summarize our findings and discuss the implications in Section ~\ref{discuss}.
The TNG simulation adopts the standard $\Lambda$CDM cosmology from \cite{Planck2016}, with parameters $\Omega_{\rm m} = 0.3089$,  $\Omega_{\rm b} = 0.0486$, $\Omega_\Lambda = 0.6911$, $H_0 = 100\,h\, {\rm km\, s^{-1}Mpc^{-1}}$ with $h=0.6774$, $\sigma_8 = 0.8159$, and $n_s = 0.9667$. 

%--------------------------------------------------------------------
\section{The Illustris TNG simulation data}
\label{data}

In this work, we use the galaxy population of the TNG300-1 simulation of the IllustrisTNG project \citep[][]{Weinberger2017,Pillepich2018b,Nelson2019,Springel2018, Naiman2018, Marinacci2018,Nelson2018}.
The TNG simulations were performed using the {\sc arepo} moving-mesh code from \citet{Springel2010}. The sub-grid physic models include a wide variety of physical processes such as star formation, stellar and supermassive black hole feedback, chemical enrichment from supernovae type II, Ia, and AGB stars, and radiative metal cooling.   

The TNG300-1 (hereafter TNG300) is the largest box of the IllustrisTNG suite with the highest resolution available for this box size with a side length of 205~$h^{-1}$~Mpc. The dark matter particle mass is of $5.9\times10^{7}$~M$_\odot$ and initial gas cells $1.1\times10^7$~M$_\odot$. In this simulation, the dark matter haloes are identified using a Friends-of-Friends ({\sc fof}) algorithm \citep{Davis1985}, while the substructures defined as subhaloes are identified using the {\sc subfind} algorithm \citep{Springel2001,Dolag2009}.

The IllustrisTNG suite has shown good agreement with several observational results at different redshifts ensuring to be a suitable tool to study the galaxy formation and halo-galaxy connection and, in particular, the alignments \citep[e.g.,][]{rodriguez21,Tenneti2021, Jagvaral2022,Zhang2022,Rodriguez2023}.
In particular, as mentioned previously, in \citet{Rodriguez2023} we use TNG300 to study the galaxy alignments of central galaxies at $z=0$ and compare them with observational results. In this study, we start from the same sample and track its evolution through the different snapshots. 

In this work, we use several subhalo and halo properties such as the virial mass, M$_{\rm 200}$, the stellar mass M$_{stellar}$ defined as the total mass of all stellar particles linked to a given subhalo. Following the selection done in \citet{Rodriguez2023}, we select as the total sample for each snapshot used those galaxies with a stellar mass greater than$10^{8.5}$~M$_\odot$ (these will be used as targets when we calculate the correlation function). The central galaxies, which are the focus of our study, will be those with an absolute magnitude brighter than $-21.5$ in the r$-$band at $z=0$. We trace back the halos that gave rise to these galaxies at earlier redshifts (and these are the centres of the correlation function at each $z$).

\subsection{Galaxy and dark matter halo shape}

The galaxy and halo shapes are modelled as in \citet{Rodriguez2023} through three-dimensional ellipsoids. For each analyzed snapshot (redshift), the inertia tensor is computed using information from the stellar and dark matter particles within two times the radius that encloses half of the mass of each subhalo. This choice roughly ensures that particles from other subhalos are unlikely to contribute to the tensor calculation.

The elements of the inertia tensor are computed as 
\begin{equation}
    I_{i,j} = \sum_n m_n x_i^n x_j^n,
\end{equation}

where $i,j$ corresponds to the three axes of the simulated box (i.e., $i,j= 1, 2, 3$), $m_n$ is the mass of the n-th cell/particle, and $x_{i}^n, x_{j}^n$ represent the positions of the n-th cell/particle in the direction $i$ or $j$. These particle positions are measured with respect to the centre of the subhalo they belong to, defined as the position of the particle with the minimum gravitational potential energy.

The eigenvalues of the inertia tensor are used to define the axis of each galaxy as $a=\sqrt{\lambda_a}$, $ b = \sqrt{\lambda_b}$ and $c=\sqrt{\lambda_c}$, with $a>b>c$.

\subsection{Merger trees, number of mergers and formation time of the dark matter haloes}

We made use of the subhalo merger trees computed with the {\sc sublink} algorithm \citep{Rodriguez-Gomez2015} to map the central galaxies selected at $z=0$ across cosmic time. For this, we map the main branch of each individual galaxy and select their dark matter, gas and stellar particles assigned at each snapshot to compute the galaxy and halo shapes.

on our analysis, we track the galaxy and halo properties and compute their shapes for snapshots 99, 67, 50, 40, and 33 corresponding to $z=0, 0.5, 1.0, 1.5,$ and $2.0$, respectively. 

Understanding the impact of mergers on galaxy alignment and shapes is one of the aims of this research. To achieve this, we use the Python module available in the IllustrisTNG database \footnote{An example script to calculate the number of mergers can be found at \url{https://www.tng-project.org/data/docs/scripts/}.
}. We focus on major mergers characterized by a stellar mass ratio  of the galaxies greater than $\frac{1}{5}$ between the haloes.

To explore the relationship between formation history and alignment, we look at the formation time ($z_{form}$) of the dark matter halo of central galaxies. To obtain $z_{form}$, we calculate the redshift at which the halo reaches half the total mass of the current time. We adopt this definition following previous work such as \cite{Lemson1999, Gao2005,Croton2007, Artale2018}.

\begin{figure}
\centering
\includegraphics[width=0.9\columnwidth]{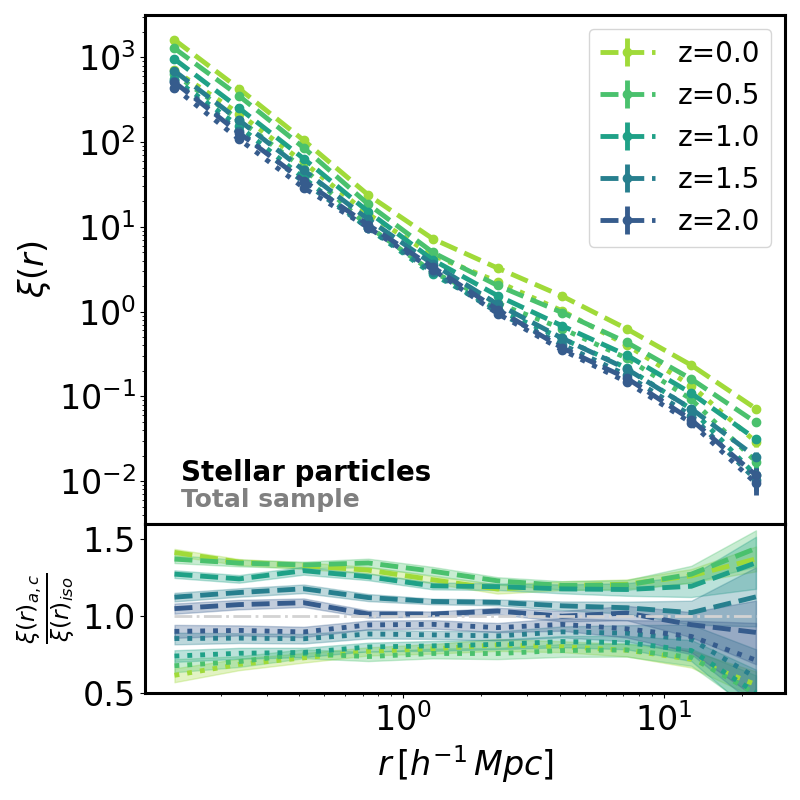}
\includegraphics[width=0.9\columnwidth]{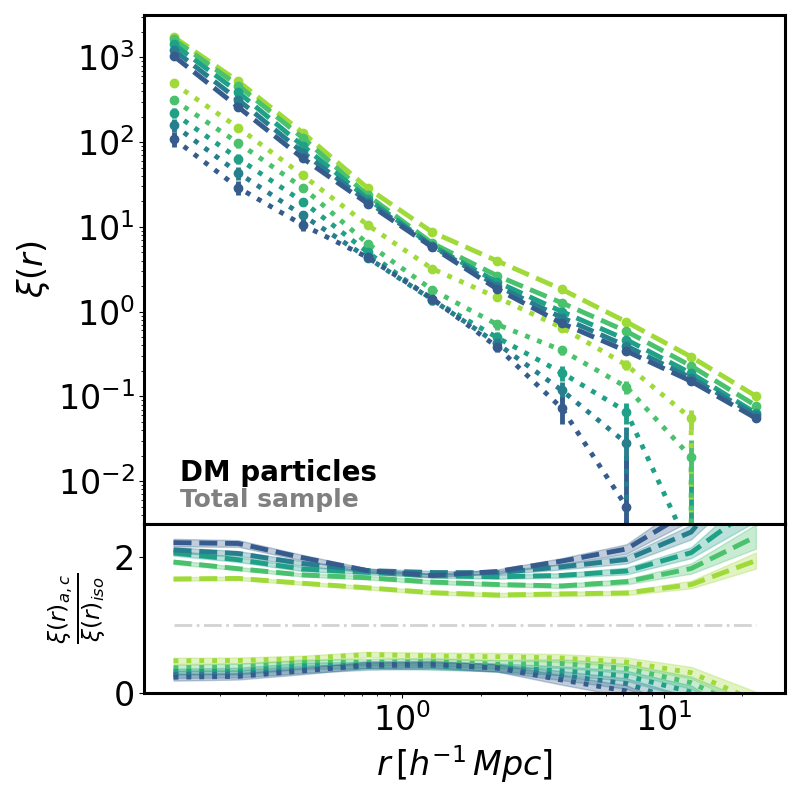}
\caption{Anisotropic correlation function at redshifts $z=0, 0.5, 1, 1.5$ and 2, for the central galaxies for the shape axes computed using stellar particles (top panel) and dark matter particles (bottom panel). The subpanels present the ratio between the correlation function along $\hat{a}$ and $\hat{c}$ axis and the isotropic correlation function (dashed and dotted lines, respectively) for the five redshifts studied. The colours aid in tracking the evolution by getting darker with increasing redshift.  
}
\label{total}%
\end{figure}

\begin{figure*}
  \centering
  \includegraphics[width=1.99\columnwidth]{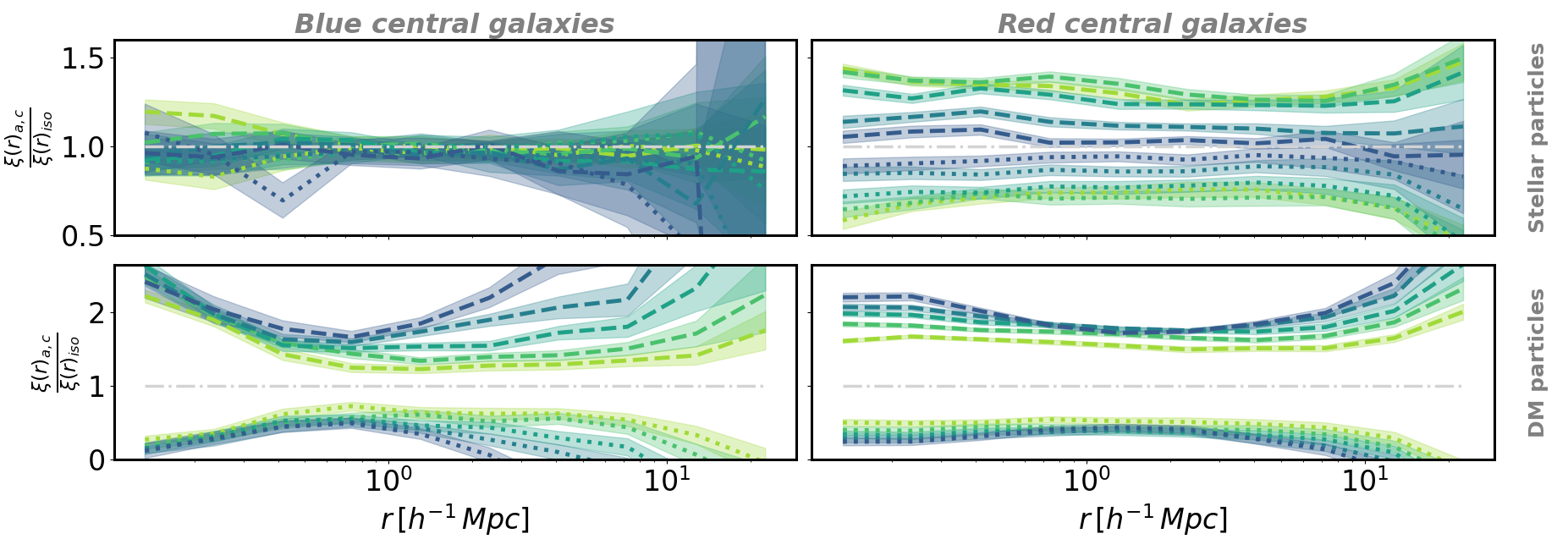}
  \caption{Evolution of the anisotropy of the correlation function for blue (left panels) and red (right panels) galaxies (identified at $z=0$) at $z=0, 0.5, 1.0, 1.5$ and 2. The top panels display the results using stellar particles, while the bottom panel shows the results using dark matter particles.
  }
  \label{colordm}%
\end{figure*}

\begin{figure*}
  \centering
  \includegraphics[width=1.98\columnwidth]{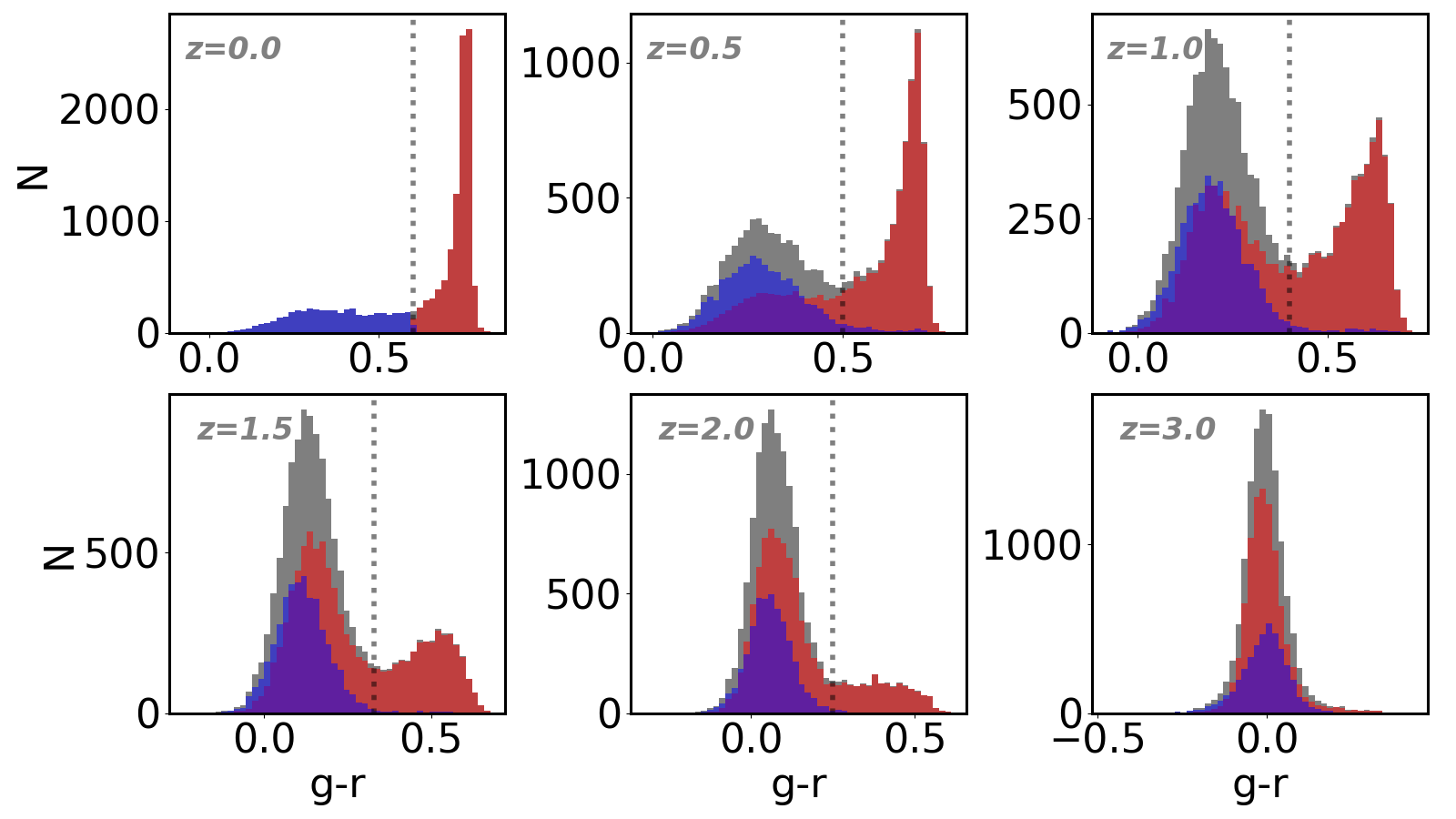}
  \caption{Evolution of the colour of galaxies. In grey, the total sample is shown, in red and blue the galaxies that had these colours at $z=0$. The vertical dotted lines represent the colour cut-off corresponding to each redshift.}
  \label{colorevo}
\end{figure*}

\begin{figure*}
  \centering
  \includegraphics[width=1.99\columnwidth]{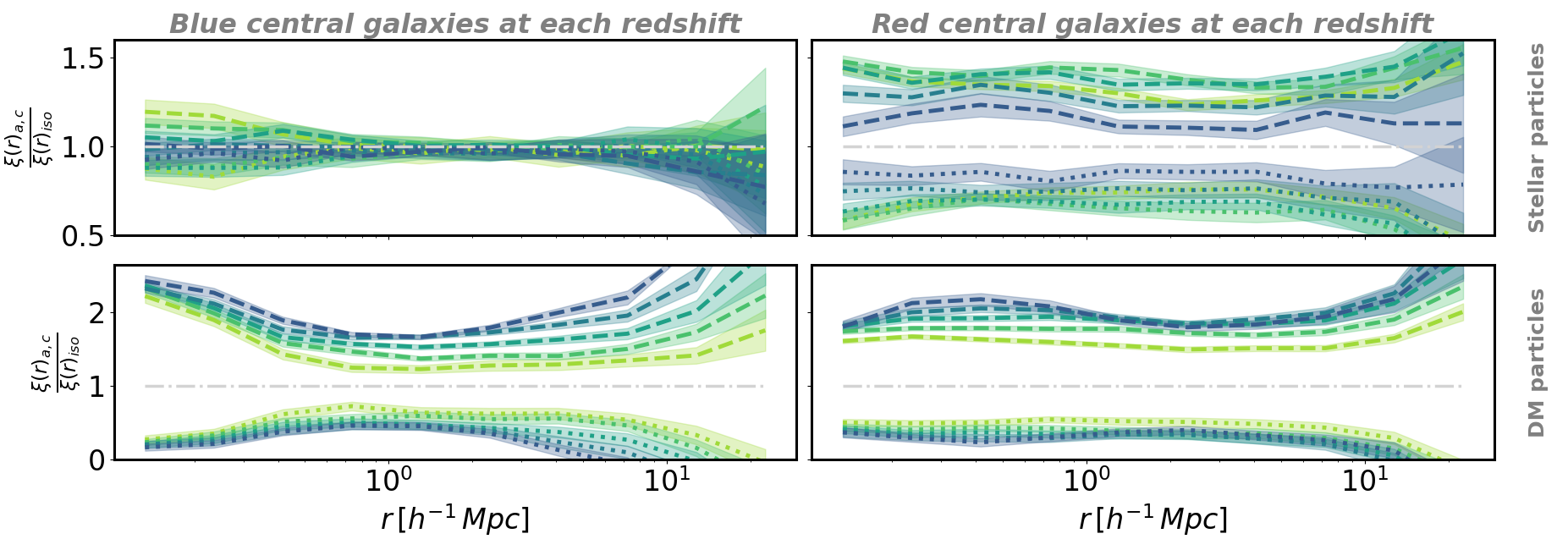}
  \caption{Evolution of the anisotropy. Following the same format as in Figure ~\ref{colordm}, but dividing the sample according to the colour cut-off corresponding to each redshift, shown in Figure ~\ref{colorevo}.
  }
  \label{colordm2}%
\end{figure*}

\section{Evolution of the anisotropic correlation function}
\label{evo}

The primary statistical tool we will use in this work is the two-point cross-correlation function ($\xi(r)$) taking the central galaxies of dark matter halos as centres and the total sample of galaxies as targets. As defined in \cite{Peebles1980}, this function measures the excess probability $dP$ of finding a target in the volume element $dV$ at a distance $r$ from a centre:
\begin{equation*}
    dP=n\, dV(1+\xi(r))
\end{equation*}
where $n$ is the target numerical density.
As in \cite{Rodriguez2022} and \cite{Rodriguez2023}, we use our own codes to calculate the correlation function adopting the \cite{Davis1983} estimator, which consists of counting central galaxy-target pairs in a distance bin, normalised to a homogeneous random distribution.

As already mentioned, in previous works, we have used a modified version of the correlation function that takes into account the excess probability in a particular direction. This is called the \textit{anisotropic correlation function}and is calculated by counting only those pairs whose positions subtend an angle less than a certain threshold with respect to a given direction (for a more illustrative picture of this procedure, see Figure 1 in \citealt{Rodriguez2023}). 
For this study, we adopt a threshold angle of 45 degrees and take the directions corresponding to the three shape axes of the central galaxy. Consequently, $\xi(r)_{a}$ denotes the anisotropic correlation function corresponding to the major axis, $\hat{a}$, $\xi(r)_{b}$ the one corresponding to the intermediate axis, $\hat{b}$, and $\xi(r)_{c}$ the one corresponding to the minor axis, $\hat{c}$. 

This work aims to study the evolution of the alignment of the bright central galaxies with their environment observed by \cite{Rodriguez2023}. Therefore, we will extend the analysis we performed in this previous work to understand the underlying nature of this phenomenon.

To begin, we calculate the anisotropic correlation function by taking as particular directions the shape axes of the central galaxies estimated using stellar and dark matter particles. Using the main branch of the merger tree for each of the central galaxies in our sample, as described in Section \ref{data}. Figure \ref{total} shows the results corresponding to five snapshots, $z=0, 0.5, 1, 1.5$ and 2. As expected, the amplitude of the correlation function increases with decreasing redshift because of the well-known evolution of the large-scale structure. This behaviour is observed both when the shape of the central galaxy is calculated using stellar (top panel) or dark matter (bottom panel) particles. The interesting point is the correlation function toward the major axis (dashed line) differs significantly from that corresponding to the minor axis (dotted line)\footnote{We do not show the intermediate axis because it lies between the other two and does not contribute to the intelligibility of the plot.}.
That is, central galaxies tend to align with the surrounding structure more in one direction than another. To quantify this difference, in the lower panel, we show the ratio between the correlation functions in the direction of each axis and the isotropic one. The alignment at $z=0$ is similar for stars and dark matter and is the one presented in \cite{Rodriguez2023}. Intuitively, this result would lead us to think that the evolution would be similar for all the components of the galaxies. However, as the figure shows, while the stars of the central galaxies align over time, their dark matter halos behave oppositely. This suggests that the physical processes driving the orientations of stars and dark matter are different. Furthermore, it is conceivable that at high redshift the stars are misaligned with their dark matter halo, so we will further explore the evolution of the alignment between these components.

Since one of the motivations of this work is to explain the dependence of alignments on colour obtained from observations, we will begin our analysis by studying the evolution of this feature.

\subsection{Color dependence evolution}
\label{color}

\begin{figure*}
  \centering
  \includegraphics[width=1.99\columnwidth]{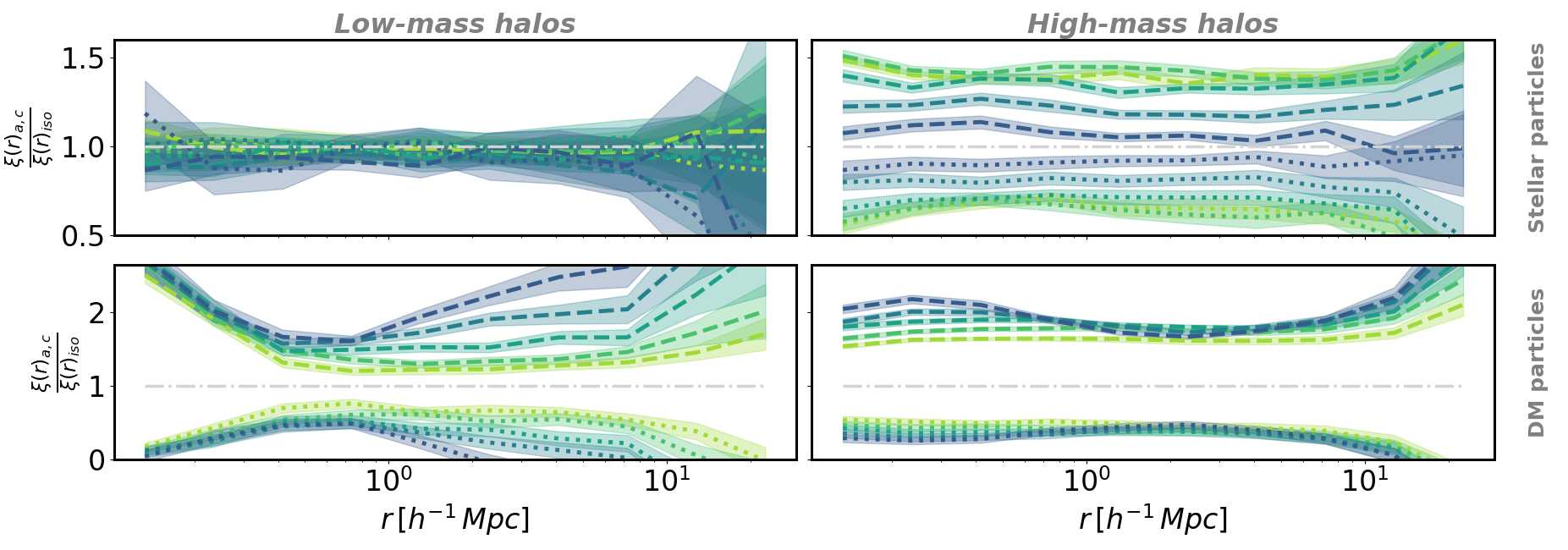}
  \caption{Following the same format as Figure ~\ref{colordm}, but splitting the sample by the total mass of the sub-halo that contains the galaxy.
  }
  \label{masadm}%
\end{figure*}

\begin{figure}
  \centering
  \includegraphics[width=0.9\columnwidth]{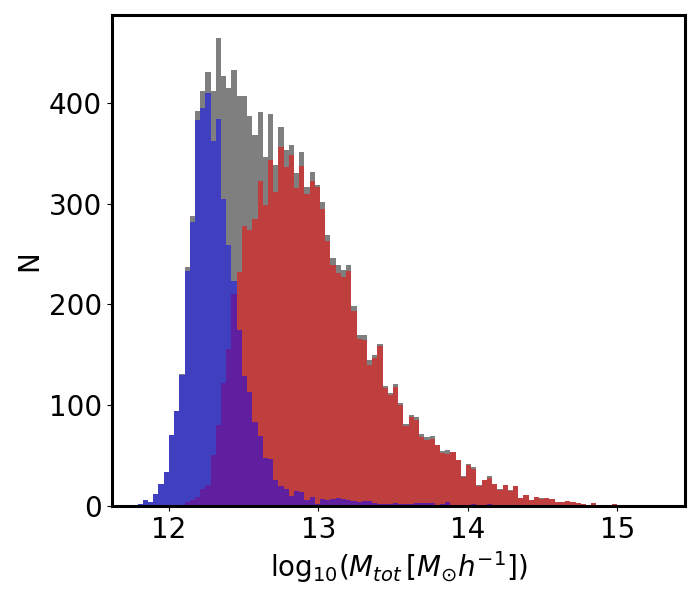} 
  \caption{{Total halo mass distribution of the full sample (grey histogram) and for the red and blue central galaxies selected at $z=0$.
   }
  }
  \label{masdist}
\end{figure}

\begin{figure*}
  \centering
  \includegraphics[width=1.99\columnwidth]{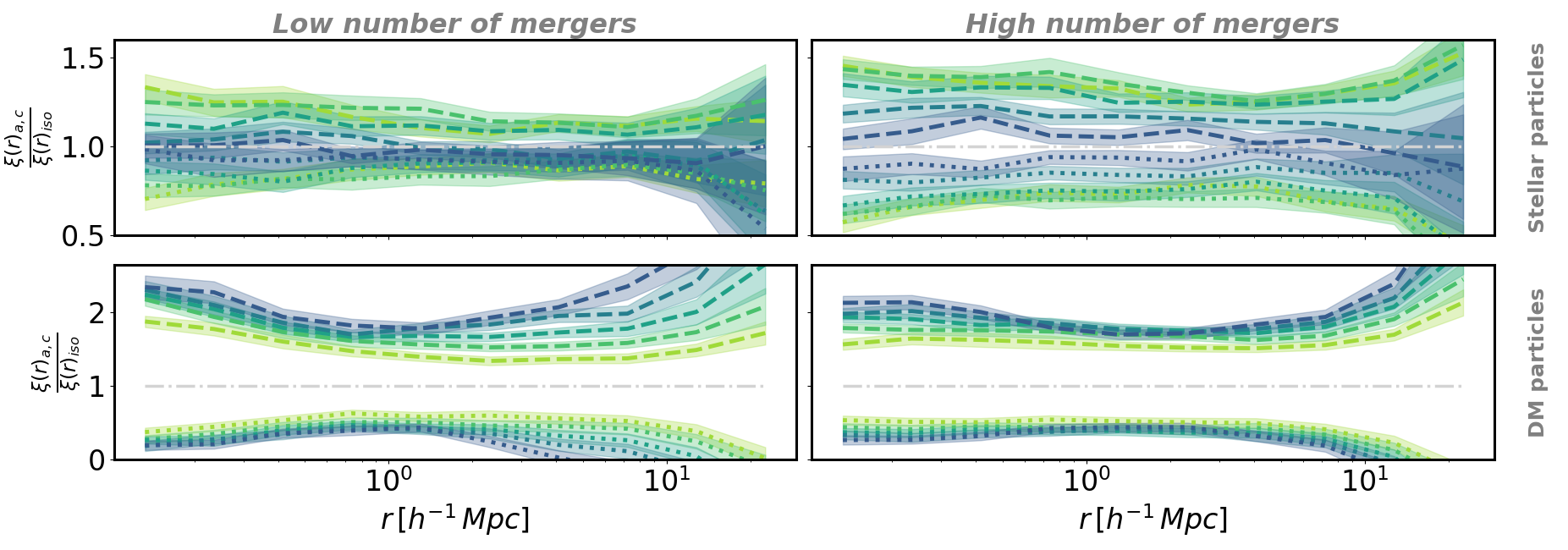}
  \caption{Evolution of the anisotropy splitting the sample by the number of mergers with 20\% or more of the central galaxy at the time of the event, following the same format as Figure \ref{colordm}.}
  \label{nmergers}
\end{figure*}

\begin{figure}
  \centering
  \includegraphics[width=0.9\columnwidth]{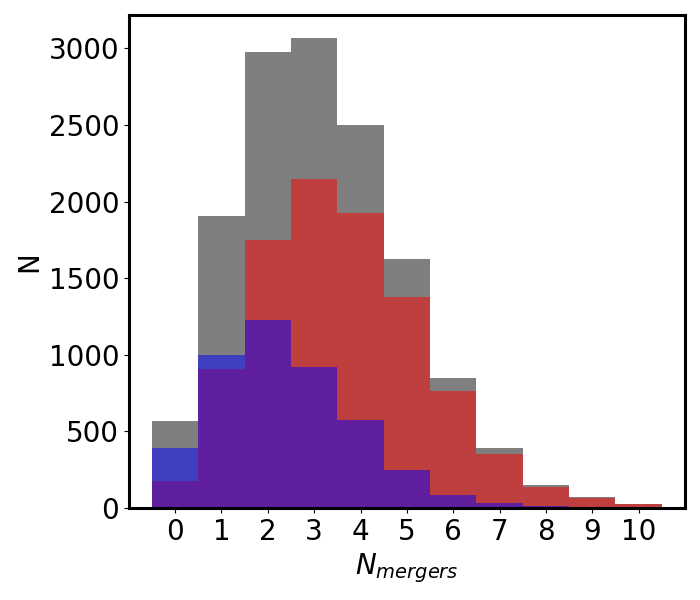} 
  \caption{Number of mergers distribution of the full sample (grey) and for the red and blue central galaxies. }
  \label{distnmergers}%
\end{figure}

\begin{figure*}
  \centering
  \includegraphics[width=1.99\columnwidth]{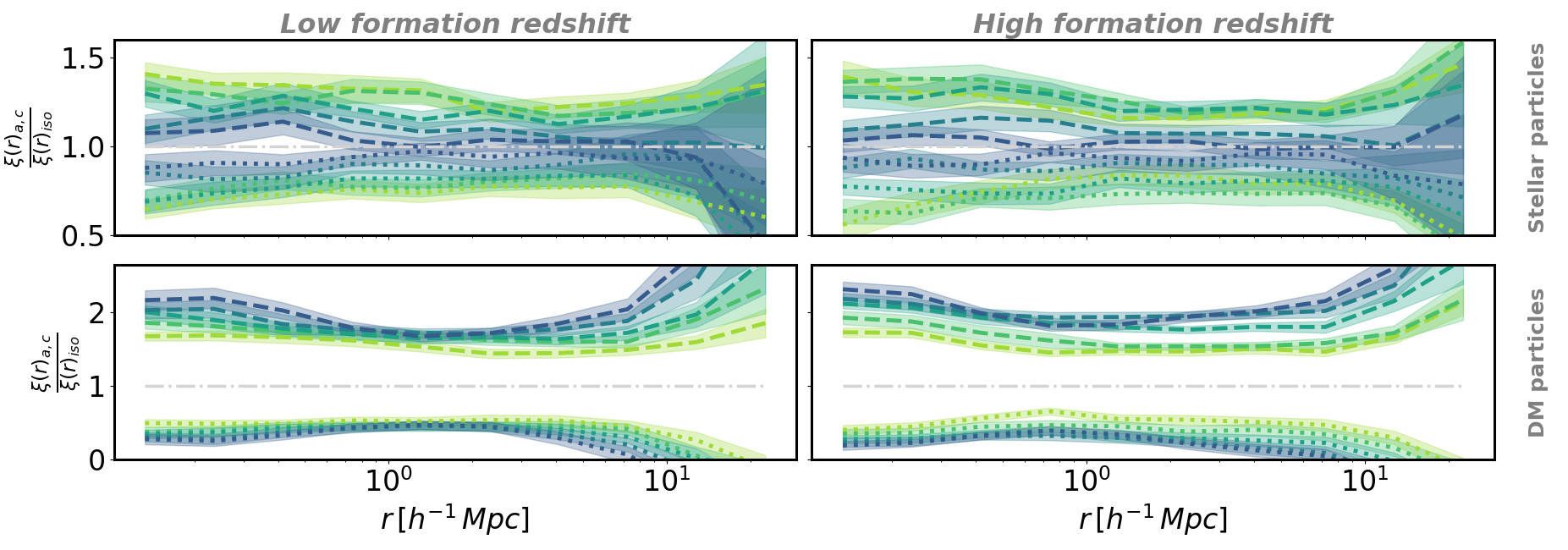}
  \caption{Evolution of the anisotropy splitting the sample by the formation time, following the same format as Figure \ref{colordm}.}
  \label{zform}
\end{figure*}

\begin{figure}
  \centering
  \includegraphics[width=0.9\columnwidth]{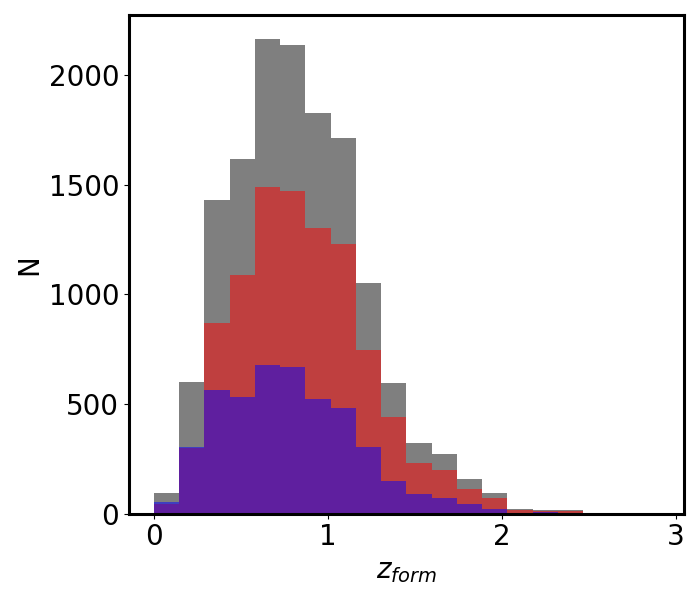} 
  \caption{Distribution of formation time, expressed in terms of redshift, for the full sample (grey), as well as for red and blue central galaxies.}
  \label{distzform}%
\end{figure}

\begin{figure*}
  \centering
  \includegraphics[width=1.5\columnwidth]{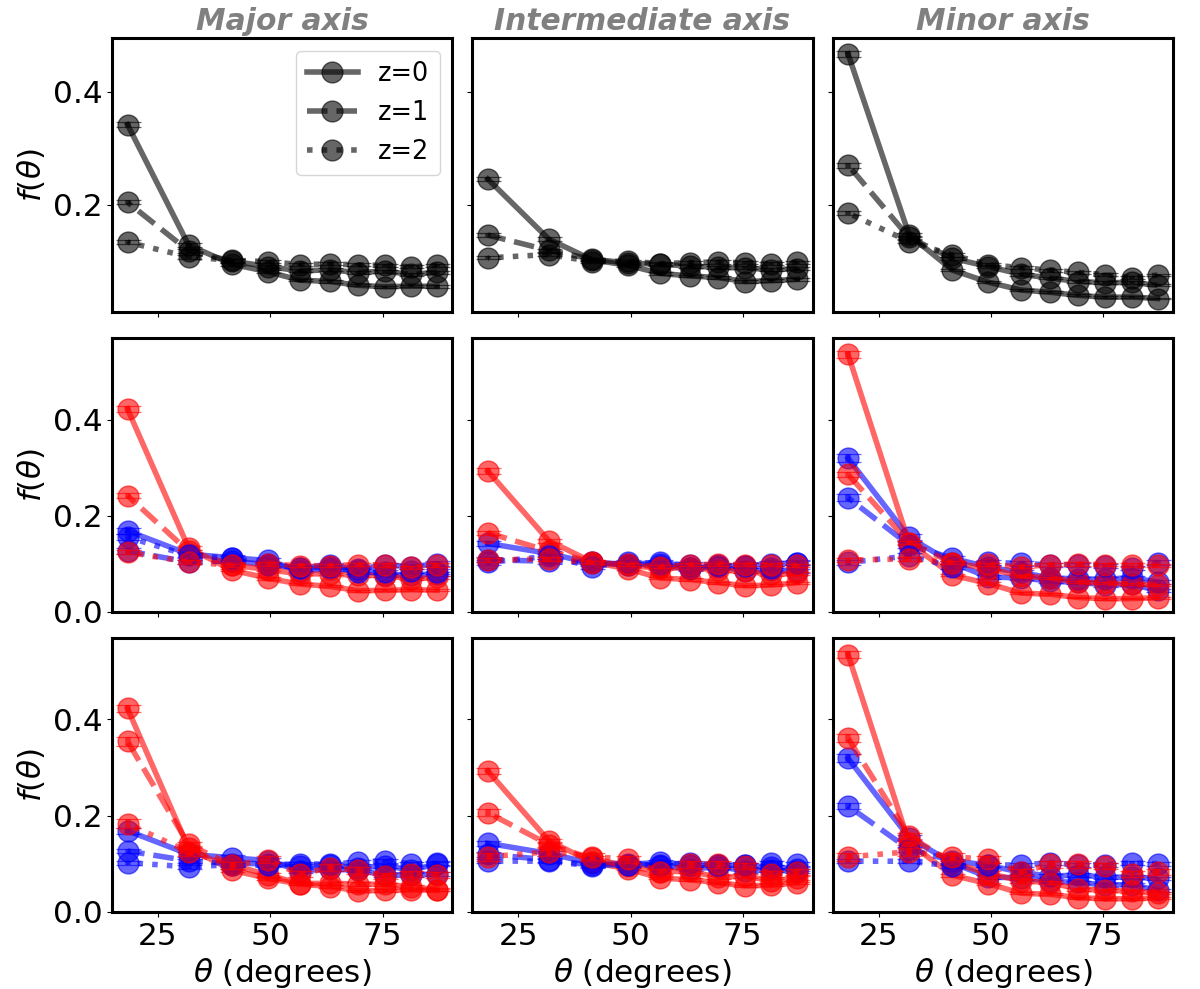} 
  \caption{The evolution of measured probability density distribution of the alignment between the dark matter and stars of the central galaxy for each of the shape tensor axes. From left to right, the major, intermediate, and minor axes are illustrated. The top panels show the results for the total sample, while the middle panels present those specific to the red and blue galaxies at z=0. Finally, the bottom panels show the findings for galaxies divided by colour with the corresponding cut-off for each redshift. Line types indicate redshifts: continuous for $z=0$, dashed for $z=1$, and dotted for $z=2$.
}
  \label{angulostardm}%
\end{figure*}

\begin{figure*}
  \centering
  \includegraphics[width=1.5\columnwidth]{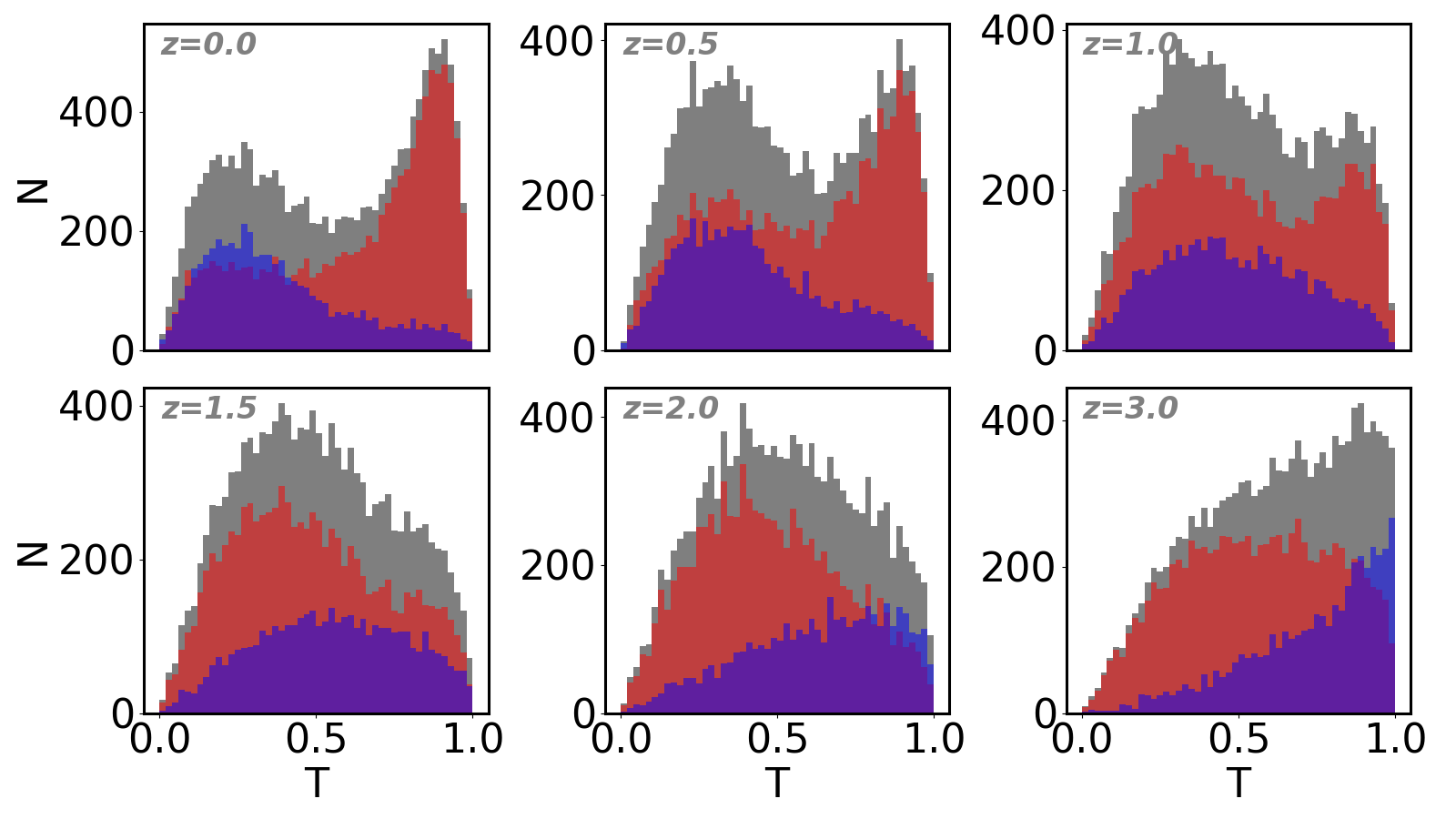}

  \caption{Evolution of triaxiality calculated with stellar particles. The entire sample is shown in grey, with red and blue galaxies represented by distinct colours. From left to right, redshifts 0, 1 and 2 are presented. }
\label{FigTriax}%
\end{figure*}

\begin{figure}
  \centering
  \includegraphics[width=0.9\columnwidth]{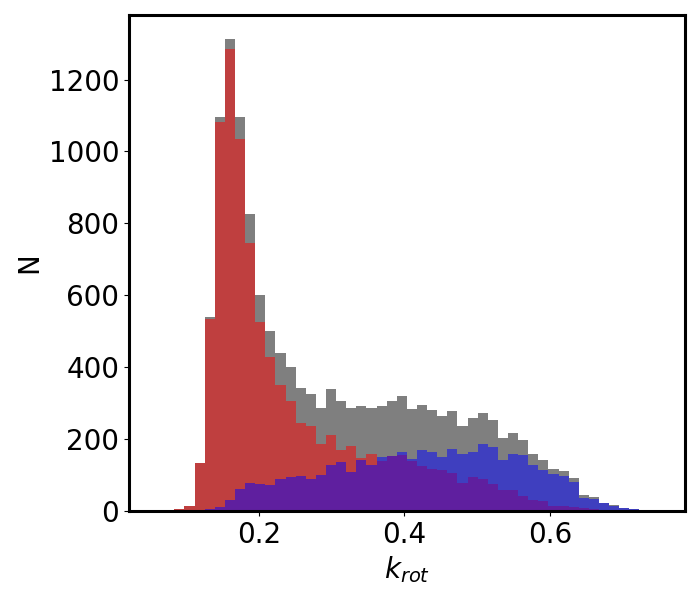} 
  \caption{ $k_{rot}$ distribution of the full sample (grey) and for the red and blue central galaxies at $z=0$.}
  \label{distkrot}%
\end{figure}

Since in the previous section we saw that the alignment of the stellar components increases with redshift, we decided to study this dependence on the colour of the central galaxy. We chose the same criterion as the one used in \cite{Rodriguez2023}, i.e. a colour cut-off of g-r=0.6 for central galaxies at $z=0$. Following the same process as in the previous section, Figure \ref{colordm} displays the evolution of the anisotropy of the correlation function for red and blue central galaxies computed using the stellar and dark matter particles (top and bottom panels, respectively). From now on, we will show only the ratios of the correlation function (in the same way as the sub-panels of Figure~\ref{total}) because it contains the relevant information about the underlying behaviour we are studying.
As shown in \cite{Rodriguez2023}, red central galaxies are responsible for the alignment signal at $z=0$. If we now analyze their evolution, we observe that the stellar component of blue galaxies never aligns, whereas red galaxies gradually align over time.

The dark matter component behaves very differently. At z=0, the alignment of red and blue central galaxies is qualitatively similar, but their evolution is not. As can be seen in Figure \ref{colordm}, although both red and blue galaxies misalign over time, reaching values similar to z=0, the blue galaxies start with higher alignment values and end up somewhat less aligned than the red ones.

By comparing the alignment of the dark matter and stellar component of blue galaxies, we can speculate that stars are misaligned with their halo. On the other hand, red galaxies exhibit a peculiar behaviour: stellar components align with their environment, while their halos misalign over time. This result reveals that the stellar component and dark matter in red and blue galaxies are influenced by different processes.
Following previous work, given that colour at z=0 is an observable property, we chose it as the initial property for our analysis. However, if the processes that determine the colour drive the alignment, then it should be the same for every redshift. It would be interesting to select galaxies at each redshift by colour to see how alignment behaves. To analyse this issue, we show in Figure \ref{colorevo} the colour distributions of our sample of central galaxies for the different redshifts. Analysing this plot, we see that the red galaxies today were blue at some point and that the division between red and blue is shifting. Setting the cut-offs so that the bimodal colour distributions characterising the galaxies provide us with two populations, we obtain colour cuts of $g-r=0.6$ at $z=0$, $0.5$ at $z=0.5$, $0.4$ at $z=1$, $0.33$ at $z=1.5$ and $0.25$ at $z=2$ (vertical lines in the figure). The results of the evolution of the alignment with these thresholds are shown in Figure ~\ref{colordm2}.
If we now compare the results of Figure~\ref{colordm} and ~\ref{colordm2} we find that the anisotropy signal is slightly larger when we split the galaxy sample by colour at a given redshift. This result implies that red galaxies that were blue at an earlier redshift did not contribute to the alignment signal.
Our findings indicate that environmental processes play a significant role in galaxy alignment among the red galaxy population, whereas this effect is not observed in the blue galaxy population suggesting that the process that reddens the galaxies aligns them. 

Beyond the previous analysis, there might be other additional and more fundamental properties that might explain our findings. 
These differences could be related to their assembly history associated with the mass and tidal fields, the mergers they underwent, and the time they formed. We will address these issues in the next section.

Previous works study the intrinsic alignments based on the galaxy morphology of IllustrisTNG and other cosmological simulations \citep{Zjupa2020}. Their findings suggest that elliptical galaxies present an increase in galaxy alignment with mass and redshift. This result is consistent with our finding for our calculation using the stellar particles in high-mass halos. Similar results are found by \citet{Yao2020}  using the DECaLS DR3 shear + photo-z catalog.

\subsection{Mass and assembly history dependence}
\label{Massandz}

It is interesting to examine how the patterns observed in the previous section for blue and red galaxy samples are exhibited when we specifically choose the galaxy samples based on their mass. 
For this purpose, we took the first and third terciles of the mass distribution at redshift $z=0$ and, following the same procedure as in the previous section, we calculated the anisotropy for each of the samples, as shown in Figure~\ref{masadm}. The results are remarkably similar to those when we divide the galaxy sample by colour. This is expected since galaxies tend to have a close relationship between colour and mass. In Figure \ref{masdist}, we present the total halo mass distribution for the complete sample, as well as for each colour sample.
Blue galaxies, with a mean mass of $2.13 \times 10^{12} M_{\odot} h^{-1}$, fill the lower end of the sample, while red galaxies, with a mean mass of $1.\times 10^{13} M_{\odot} h^{-1}$, dominate the upper end. 
Our results suggest that mass and colour can be used interchangeably in our alignment analysis and that colour is a good observational proxy for mass\footnote{Note that the division between red and blue is made taking into account the typical bimodal colour distribution, and together cover the total sample while the mass samples correspond to the outer terciles. We did this to be consistent with the other analyses, but the result is the same if we choose them according to the median mass.}.

Previous reports suggest that the alignment processes are linked to galaxy mergers producing modifications in both the stellar component and the dark matter \citep{RagoneFigueroa2020,Lee2022}. To examine how these events affect the alignment of central galaxies, we divided the sample of galaxies according to the number of 
mergers. For each central galaxy, we compute  the number of merger events computed as those where the stellar mass ratio is above of 1/5.
We select two samples corresponding to the first and third tercile of the distribution of the number of mergers.
We present our results in Figure~\ref{nmergers}. 
We find that the anisotropy for the stellar component shows a similar evolution for low and high numbers of mergers, being stronger for the latter.
On the other hand, we find that the anisotropy using dark matter particles presents an evolution opposite to that found for stellar particles for low and high numbers of mergers. This trend is in line with our findings when we split the galaxy sample by halo mass and colour.

Our results suggest that galaxies with a high number of mergers present a mild correlation between the galaxy alignment and the environment in which they live.
This could be explained because galaxies with a low number of mergers are a mixture of blue and red galaxies, as illustrated in the histogram of Figure \ref{distnmergers}, with the red ones being responsible for the increase in the alignment signal.

The last property we will analyze in this section is the formation time described in Section \ref{data}. We consider that the time at which each halo was assembled could be related to the alignment with the surrounding structure. Whatever process led to the halos being aligned or misaligned today, it should have acted longer on the older halos. Following the usual procedure used up to this point, we select two samples corresponding to the first and third terciles according to the time of formation. In Figure \ref{zform}, we show the alignment for each of these samples. Regardless of whether we calculate the shape with the dark matter or the stellar component, both young and old halos statistically have similar orientations.
This result contradicts the hypothesis that formation time could be a decisive factor in the alignment of central galaxies. Based on the histogram displayed in Figure \ref{distzform}, it can be observed that the formation time follows a similar pattern for blue and red galaxies, further supporting the notion that colour and mass are the properties most strongly associated with alignment.

\subsection{Stellar-dark matter alignment}

In previous sections, we study the anisotropy and alignments between the different components of the central galaxies and their surroundings.
Using the stellar particles, we find that red galaxies start misaligned at high redshift and end up aligned at low redshift. This trend reverts when we use dark matter particles.
Blue galaxies, instead, 
present misalignment for the stellar particles in all the redshifts studied, while using dark matter particles we find alignment at high redshift that vanishes as redshift decreases.
So, we can infer that the evolution of stars and dark matter is quite different. This may be because of changes in the orientations between the stellar component and their dark matter halos.
To further enhance the characterisation of the surrounding environment, we will incorporate information on internal scales by estimating the alignment of stars relative to their dark matter haloes. This will provide us with another valuable insight in addition to the correlation function analysis.

In order to measure how the different parts of the central galaxies are lined up, we calculate the angle between the axes of the shape tensor using both stars and dark matter. Then, we make the probability distribution of this angle, i.e. the ratio between the number of galaxies with a given angle and the total. This will help us track how the alignment between these components changes over time. Figure ~\ref{angulostardm} shows the distribution of the three main axes of shape for red and blue central galaxies at different redshift values ($z=0.0$, $1.0$ and $2.0$). It can be observed that in general, dark matter and stars align along their three axes over time. When analysing the behaviour of the different axes for the total sample, it is the minor axis that shows a significantly stronger alignment (top right panel), followed by the major axis (top left panel). When the sample is categorized into red and blue central galaxies at $z=0$ as before (middle panels), it is observed that regardless of the redshift value, red galaxies contribute the most to the alignment signal. Blue galaxies, in turn, present a very weak alignment signal in all redshifts, except the minor axis case, where an evolution with redshift is observed but is not as strong as that of the red galaxies. This is to be expected, given that the minor axis is closely aligned with the angular momentum that defines the morphology of spiral galaxies. To examine whether the processes that redden the galaxies are the same as those that produce the internal alignment of the stars with their own halo, we repeat the procedure described in section \ref{color}, selecting by colour at each redshift. The behaviour is very similar to those obtained by dividing between red and blue at $z=0$ (obviously the same for $z=0$), the main difference being that the alignment signal is stronger for the intermediate redshift ($z=1$). This reinforces the idea that the processes that redden galaxies are the ones that align them not only with the environment but also internally.

So far, we find that the alignment between dark matter and stars increases with time and is smaller for blue galaxies than for red ones. On the other hand, blue galaxies tend to be flattened structures while red galaxies tend to be spheroids. Thus, the shape of both the stars and the dark matter could be a critical aspect of the phenomenon under examination. To study the evolution of the central galaxy shapes we choose the following triaxiality ($T$) parameter \citep{Franx1991,Warren1992}:
\begin{equation}
    T=\frac{a^2-b^2}{a^2-c^2}
\end{equation}
where $a$, $b$ and $c$ are the axes of each galaxy defined from the eigenvalues of inertia as described in section \ref{data}.  We say that a halo has a prolate shape when T$\sim 1$, while an oblate shape is defined as T$\sim 0$.

Our findings point out that the distribution of the T-parameter for dark matter is very similar regardless of the colour of the galaxies that inhabit them. They evolve from slightly prolate shapes towards triaxial structures. 
In contrast, when we calculate the shape using the stars, the differences in triaxiality become more interesting. Figure \ref{FigTriax} shows these distributions for redshift between 3 and 0 (we include $z=3$ to follow the story a bit further back). The T values of the blue galaxies, often associated with spirals, start quite prolate at $z=3$ and become more oblate with time.
When we focus on the T distribution for red central galaxies, we find that tend to move from oblate at $z=3$ to prolate at $z=0$.

The changes in the shape of galaxies may be due to an evolutionary process. Less massive blue galaxies are less evolved compared to more massive red galaxies. During the period studied, the processes that form disk-like structures in blue galaxies are more efficient. According to widely accepted models, baryons start with a similar T distribution as their halo and star-forming gas cools and collapse into high-density regions acquiring high rotation. On the other hand, red galaxies at a redshift of z=2 might have already passed that stage. They start having flattened structures at this redshift (upper right panel of Figure 1), which indicates that they are in a more advanced evolutionary stage and will be transformed into spheroids by a mixture of complex astrophysical processes dominated by mergers.
This is consistent with \cite{Tacchella2019}, which is summarized in Figure 14 therein. 

The shape evolution should also be reflected in the colour of the galaxies, since, as we know, disc galaxies tend to be blue and spheroidal galaxies red. Figure \ref{colorevo} shows how the colour of red and blue galaxies (selected at z=0) evolves. Unlike the shape, the two selected populations do not differ in colour at high redshift (z=3), which indicates that although the red galaxies are in a more evolved dynamical stage than the blue ones, their stellar component has not yet had time to evolve enough to differentiate from the blue ones. As time goes by, red galaxies (more massive) undergo more complex physical processes (mergers, close interactions, gas depletion, chemical enrichment, etc.) that differentiate the behaviour of their stellar component, making them redder as time passes.

Disc-like galaxies are supported by high rotation, whereas spheroidal galaxies are dominated by velocity dispersion. Consequently, a useful way to characterize galaxies is using stars to determine the fraction of kinetic energy associated with ordered rotation with respect to the total kinetic energy \citep[$k_{rot}$, see equation 1 in ][]{Sales2012}. This parameter takes values equal to 1 for discs with perfect circular motions and tends to zero for systems supported by velocity dispersion. In Figure \ref{distkrot}, we display the distribution of this parameter for our sample. We superimpose the distributions for galaxies that are red and blue at $z=0$, in order to examine the relationship between colour and dynamics. We can observe that the dynamic characteristics of the galaxies correspond well with their colour selection. Therefore, when selecting galaxies based on this parameter, the alignments found are very similar to those obtained when selecting based on colour. This indicates that, like mass, colour is a good proxy for dynamics.

\section{Summary and conclusions}
\label{discuss}

This research is based on our previous findings, where we examined how central galaxies align with their surroundings. In \cite{Rodriguez2022}, we used the SDSS galaxy groups and found a significant dependence between the alignment and the colour of central galaxies. Specifically, we observed that red central galaxies are more aligned with their environment than blue ones. This pattern was seen for both alignment with galaxies in the same group and with the environment on larger scales. Additionally, we noticed that the major axis of the red central galaxies is highly aligned with the major axis of their respective group. To explain this phenomenon, we performed an analogous analysis in \cite{Rodriguez2023}, but this time using the TNG300 simulation. We found the same dependence of the alignment of the central galaxies on colour and mass. However, the latter is not detected in observations.
One notable result was that the stars in the red central galaxies are strongly aligned with their dark matter halos, while the blue ones are not. However, the dark matter halos of all the central galaxies show strong alignment with the host halo. All these results suggest that the alignments are linked at different scales, possibly involving different physical processes.
Our previous results are in agreement with previous works using simulations \citep[e.g.,][]{Zjupa2020,Tenneti2021,Delgado2023} and observations \citep[e.g.,][]{Sales2004,yang2006alignment,Huang2016,Huang2018}.

The results summarized above describe the behaviour of the orientation of central galaxies with their surroundings in both observations and simulations at present. In this work, we build on our previous findings and study in detail how these alignments develop over time. In particular, we study the alignment of the stellar- and dark matter components.

We initially focused on the alignment with the surrounding structures. For this purpose, we use the anisotropic correlation function in the major and minor axis directions of the shape tensor calculated using the stellar and dark matter particles separately. For the total sample of central galaxies analysed, we observe that as time passes, the stars become more aligned with the environment, whereas the opposite happens when we calculate the shape tensor using the dark matter particles.
Based on the similar alignment of stars and dark matter at $z=0$, it might be tempting to assume that all components of galaxies are aligning together.
However, the observed evolution (Fig. \ref{total}) contradicts this assumption as stars and dark matter evolve in opposite directions, suggesting that different evolutionary processes determine alignment.

As we already know, there is a dependence between alignment and colour.
In this work, we analyse the evolution of this dependence in time.
At a glance, we find that the evolution of blue and red galaxies is quite different for both their stellar and dark matter components. 

We find that the anisotropic correlation of red galaxies presents a similar evolution to the total sample for both the stellar component and the dark matter (right-hand panels of Fig. \ref{colordm}). 
Conversely, the alignment evolution of blue galaxies is very different.
Their stellar component shows no signs of evolution throughout their history, i.e. they are always misaligned (upper left panel of Fig. \ref{colordm} ). However, their dark matter halo shows an evolution from a high alignment at the earliest times to an alignment similar to that of the total sample at present (lower left panel of Fig. \ref{colordm}). To investigate if there is a connection between the alignment phenomenon and the physical processes that caused the galaxies to become redder, we looked at their colour evolution (see Fig. \ref{colorevo}).
For this, we selected galaxies based on their colour distribution at different redshifts and computed their anisotropy signal. By comparing Fig.~\ref{colordm} and ~\ref{colordm2} our results suggest that red galaxies (or the process that reddens them) are the main drivers for galaxy alignment.

Since colour correlates strongly with mass and $k_{rot}$ (Figs. \ref{masdist} and \ref{distkrot}), it is natural that the results obtained when selecting for mass or $k_{rot}$ were also very similar to those for colour. This confirms that colour is an excellent proxy for mass and dynamics, at least as far as alignment is concerned.

To understand the origin of this alignment, one can think that the number of mergers and the formation time of the galaxies are quantities that could account for the evolutionary process taking place in the galaxies. 
However, the alignment with the environment beyond the one-halo term does not seem to depend strongly on these variables, as we only find a weak dependence on the number of mergers and almost negligible dependence on the formation times. 

It became evident that studying the internal behaviour of central galaxies is necessary after observing differences in the alignment behaviour of the dark matter halo and the stellar component. To accomplish this, we used the probability distribution of the angle between the axes of the shape tensor computed for the stellar and dark matter components.
This allowed us to connect the internal evolution of one component to the other. Our findings indicate that, on average, the alignment between stars and dark matter increases with time in the total sample (upper panels of Fig. \ref{angulostardm}). When we categorized the galaxies into red and blue, we noticed that stars in red galaxies align more strongly
over time with their dark matter halo than the blue ones.
This trend becomes more pronounced when choosing the red galaxy sample at a specific redshift (see bottom panel of Fig.~\ref{angulostardm}) compared to selecting the red galaxy sample at $z=0$ (middle panel of Fig.~\ref{angulostardm}). 
These findings imply that the environmental mechanisms that transform galaxies into red/quenched have a significant correlation with their alignment.
In addition, we find distinct evolution patterns in the shape tensor of red and blue galaxies (Fig. \ref{FigTriax}), with red galaxies evolving into prolate galaxies and blue galaxies evolving into oblate shapes. 

The aforementioned results confirm that alignment is a dynamic phenomenon that evolves due to different processes acting at different scales. 
Although the alignment of dark matter halos at $z=0$ is strong on large scales, they arrived at this state after a misalignment process, which may be a consequence of non-linear processes after gravitational decoupling from their surroundings.  Consequently, the progressive alignment of stars on large scales is a manifestation of the alignment with the final state of their halo and is not necessarily related to the distribution of matter on large scales.
This internal alignment occurs predominantly for red galaxies (i.e. the most massive ones), which would be explained by their assembly history. The fact that the results obtained do not depend on whether these galaxies are selected (at z=0 or at each redshift) seems to indicate that there is a relationship between the quenching and the alignment process.
Although the formation time is very similar for red and blue galaxies (Fig. \ref{distzform}), the average number of mergers for red galaxies is $\sim 4$, while for blue galaxies it is $\sim 2$. Thus, mergers seem to be the main responsible for the alignment of stars with their halos. This process also impacts the stellar component, making up the shape and color of the central galaxies observed today, causing stars in the red galaxies to progressively align with their haloes and mutate from oblate to prolate. In this scenario, at high redshift, the central galaxies do not differentiate in colour but in mass (at z=2, $1.6\times10^{12}$~M$_\odot$ and $4.0\times10^{11}$~M$_\odot$ for red and blue galaxies respectively). Later, the mergers acting on the more massive ones cause them to redden and become oblate, more aligned with their halo and, therefore, with their surroundings. The initial difference in mass between those that are currently red and blue would also explain the difference in the initial alignment of the dark matter component of these with the surroundings.  It would be interesting to address this better in future work.

In summary, the mergers modify the red central galaxies morphology and also cause alignment. In contrast, the blue ones are younger and have recently acquired their oblate shape, undergone fewer mergers, and are less massive, resulting in less active alignment processes. 
From the inside out,  the alignment occurs on different scales. The stellar component of the central galaxies aligns with the dark matter of their halo as they evolve morphologically throughout mergers. The halos of these galaxies, in turn, were aligned with the environment in the past, but they are gradually losing this alignment, although they remain relatively strong even today.

Based on our analysis of previous findings, we have identified areas that need further investigation in future works. In the simulation, we could look at some specific features, such as the effects of mergers on alignment and the influence of factors such as impact parameters, directions, and mass ratios. Additionally, we need to gain a better understanding of the alignment of each axis and angular momentum, particularly in the early stages, and its relationship with the surrounding environment. It would also be interesting to compare and assess our results with observations. 
We plan to further investigate on this aspect in the future.
    
\begin{acknowledgements}
      The authors wish to thank the anonymous referee for his/her thorough report that helps us to improve this manuscript.
      FR and MM thanks the support by Agencia Nacional de Promoci\'on Cient\'ifica y Tecnol\'ogica, the Consejo Nacional de Investigaciones Cient\'{\i}ficas y T\'ecnicas (CONICET, Argentina) and the Secretar\'{\i}a de Ciencia y Tecnolog\'{\i}a de la Universidad Nacional de C\'ordoba (SeCyT-UNC, Argentina).
      FR would like to acknowledge support from the ICTP through the Junior Associates Programme 2023-2028. MCA acknowledges support from ANID BASAL project FB210003.
\end{acknowledgements}

% for the bibliography, at the end
\bibliographystyle{aa} % style aa.bst
\bibliography{main} % your references Yourfile.bib
\end{document}